\documentclass[aps,nofootinbib,longbibliography,prd,twocolumn]{revtex4-1}
\usepackage[babel]{csquotes}
\usepackage{graphicx}
\usepackage{amsmath,amssymb}
\usepackage[colorlinks,citecolor=blue,linkcolor=blue,urlcolor=blue]{hyperref}
\usepackage{mathrsfs}
\usepackage{enumerate}
\usepackage{dsfont}
\usepackage{mathrsfs}
\def\be{\begin{equation}}
\def\ee{\end{equation}}

\usepackage{verbatim} 

\usepackage{scrextend} 
\usepackage{float}

\begin{document}

\title{Volume Entropy}
\author{Valerio Astuti\thanks{valerio.astuti@roma1.infn.it}, Marios Christodoulou\thanks{christod.marios@gmail.com}, Carlo Rovelli\thanks{rovelli@cpt.univ-mrs.fr}}

\affiliation{\small
\mbox{CPT, Aix-Marseille Universit\'e, Universit\'e de Toulon, CNRS, F-13288 Marseille, France.} }
\date{\small\today}

\begin{abstract}
\noindent Building on a technical result by Brunnemann and Rideout on the spectrum of the Volume operator in Loop Quantum Gravity, we show that the dimension of the space of the quadrivalent, diffeomorphism invariant states with no zero-volume nodes describing a region with total volume smaller than $V$ has \emph{finite} dimension, bounded by $V \log V$.   This allows us to introduce a notion of  ``volume entropy" for this phase space: the von Neumann entropy associated to the measurement of volume.

\end{abstract}

\maketitle 

\section{Introduction}

Bekenstein's introduction of an entropy associated to the horizon of a black hole \cite{BekensteinBlackHolesEntropy1973a} was a first indication of the existence of thermodynamical aspects in the dynamics of spacetime.  This result led to the formulation of the ``laws of black holes thermodynamics" by Bardeen, Carter, and Hawking \cite{Bardeenfourlawsblack1973} and to Hawking's discovery of black role radiance, which reinforced the geometry/thermodynamics relation  \cite{HawkingParticlecreationblack1976}.  The connection between Area and Entropy indicates that aspects of space-time can be treated statistically at scales large compared to the Planck length \cite{JacobsonThermodynamicsSpacetimeEinstein1995}; this is true whether or not the microscopic degrees of freedom of gravity are simply its quanta \cite{ChircoSpacetimethermodynamicshidden2014} as assumed in Loop Quantum Gravity (LQG), or else.  In particular, black hole entropy can be interpreted as cross-horizon entanglement entropy (see \cite{BianchiEntanglement,Bianchiarchitecturespacetimegeometry2014} for recent results reinforcing this interpretation, and references therein), or --perhaps equivalently-- as the von Neumann entropy of the statistical state representing a macrostate with given horizon Area. In the context of Loop Quantum Gravity, this second possibility was considered in \cite{RovelliBlackHoleEntropy1996} and later extensively analyzed; for a recent review and full references see \cite{PerezBarbero,AbhayILQGS}.

All these developments are based on the idea of studying thermodynamical properties of spacetime \emph{surfaces}. This association has motivated the various versions of the holographic hypothesis:  the conjectural idea that the degrees of freedom of a region of space are somehow encoded in its boundary.    

In this paper, instead, we study statistical properties of quantum states of geometry associated to spatial \emph{regions}.  On the possible physical relevance of the von Neumann entropy associated to the \emph{Volume} of a region see for instance \cite{ChristodoulouHowbigblack2015a, ChristodoulouVolumeoldblack2016b,PerezNofirewallsquantum2015,Rovelli17}. Definitions of entropy associated to bulk degrees of freedom of a spin network were already tentatively explored in \cite{MeissnerEigenvaluesvolumeoperator2006,RovelliSingleparticlequantum2010a}.\footnote{In \cite{RovelliSingleparticlequantum2010a} the authors derive the BGS entropy of a graph with a particle living on it. Volume information, the main observable we want to associate an entropy with, was not considered. In addition the authors note that the source of their entropy is not a genuine density matrix associated to the gravitational field, but rather a description of the matter energy levels in a given geometry. In \cite{MeissnerEigenvaluesvolumeoperator2006} the entropy associated to the bulk degrees of freedom is suggested in the last sentence as an open possibility.} Here we show that under certain hypotheses it is possible to define a von Neumann entropy for the quantum gravitational field, associated to the Volume of a region, and that this entropy is (under suitable conditions) finite. 

To this aim, we prove a technical finiteness result on the number of quantum states of gravity describing a region of finite volume.   More precisely, we work in the context of LQG as defined in \cite{RovelliCovariantloopquantum2015, QuantumGravityRovelli},\footnote{Our results are not directly applicable to other formulations of Loop Quantum Gravity \cite{ThiemannLecturesLoopQuantum2003, ThiemannModernCanonicalQuantum2007} and would appear to present further complications, in particular by the use of an extended state space through the inclusion of graphs with higher-valent nodes and by defining diff-invariant spin-networks states through explicit embeddings in hypersurfaces. These issues, which do not arise in the formulation of \cite{RovelliCovariantloopquantum2015,QuantumGravityRovelli}, are not addressed in this article. We thank an anonymous referee for pointing this out.} and we prove that the dimension of the space of diffeomorphism invariant quadrivalent states {\em without zero-volume nodes}, describing a region of total volume smaller than $V$, is finite.  We give explicitly the upper bound of the dimension of this space as a function of $V$. The proof is based on a result on the spectrum of the LQG Volume operator proven by Brunnemann and Rideout \cite{BrunnemannPropertiesvolumeoperator2008a, BrunnemannPropertiesvolumeoperator2008b}.  Using this, we define a von Neumann entropy of a quantum state space of the gravitational field, associated to Volume measurements.

%

\section{Counting spin networks}  \label{sec:2}

Consider the measurement of the volume of a 3d spacelike region $\Sigma$. The physical system measured is the gravitational field. In the classical theory, this is given by the metric $q$ on $\Sigma$: the volume is $V=\int_\Sigma \sqrt{\det q} \ d^3x$. In the quantum context, using the LQG formalism, the geometry of $\Sigma$ is described by a diffeomorphism invariant state in the kinematical Hilbert space $\mathcal{H}_{\rm\emph{diff}}$. The volume of $\Sigma$ is described by a volume operator $\hat{V}$ on this state space. We refer to \cite{RovelliCovariantloopquantum2015,QuantumGravityRovelli} for details on basic LQG results and notation.

We restrict $\mathcal{H}_{\rm\emph{diff}}$ to four-valent graphs $\Gamma$ where the nodes $n$ have non-vanishing (unoriented) volume $v_n$.  The spin network states $|\Gamma, j_l , v_n\rangle \in \mathcal{H}_{\rm\emph{diff}}$, where $j_l$ is the link quantum number or spin, form a countable, orthonormal basis of $\mathcal{H}_{\rm\emph{diff}}$. (We disregard here eventual additional quantum numbers such as the orientation, that have no bearing on our result.)  The intertwiner basis at each node is chosen so that the local volume operator $\hat{V}_n$, acting on a single node, is diagonal and is labelled by the eigenvalues $v_n$, of the node volume operator $\hat{V}_n$ associated to the node $n$. 
\be
 \hat{V}_n \;|\Gamma , j_l, v_n\rangle   = v_n  |\Gamma , j_l, v_n\rangle
\ee
The states $|\Gamma, j_l , v_n\rangle  $ are also eigenstates of the total volume operator $\hat{V}=\sum_{n=1}^N \hat{V}_n $, where $N$ is the number of nodes in $\Gamma$, with eigenvalue  
\be
v   = \sum_{n=1}^N v_n, 
\ee
the sum of the node volume eigenvalues $v_n$.

We seek a bound on the dimension of the subspace $\mathcal{H}_V$ spanned by the states $|\Gamma, j_l , v_n\rangle$ such that $v\leq V$.
That is, we want to count the spin-networks with volume less than $V$. We do this by bounding the number $N_\Gamma$ of  four valent graphs in $\mathcal{H}_V$, the number $N_{\{j_l\}}$ of possible spin assignments, and the number of the volume quantum numbers assignments $N_{\{v_n\}}$ on each such graph. Clearly
\begin{equation} \label{eq:boundDim}
\dim \mathcal{H}_V \leq N_{\Gamma}\ N_{\{j_l\}}  N_{\{v_n\}}.
\end{equation}

Crucial to this bound is the analytical result on the existence of a volume gap in four-valent spin networks found in \cite{BrunnemannPropertiesvolumeoperator2008a,
BrunnemannPropertiesvolumeoperator2008b}. The result is the following. In a node bounded by four links with maximum spin $j_{max}$ all non-vanishing volume eigenvalues are larger than
\begin{equation} \label{eq:vmin}
v_{gap} \geq  \frac{1}{4\sqrt{2}}\; {\ell_{\rm P}^3 \gamma^{\frac{3}{2}}} \sqrt{j_{max}} 
\end{equation}
Where $ \ell_{\rm P}$ is the Planck constant and  $\gamma$ the Immirzi parameter. 
 Numerical evidence for equation \eqref{eq:vmin} was first given in
\cite{BrunnemannSimplificationspectralanalysis2006} and a compatible result was estimated in \cite{BianchiDiscretenessVolumeSpace2011a}. 
Since the minimum non-vanishing spin is $j=\frac12$, this implies that 
\begin{equation} \label{eq:vmin2}
v_{gap} \geq \frac{1}{8}\;{\ell_{\rm P}^3 \gamma^{\frac{3}{2}} }\equiv v_o
\end{equation}

From the existence of the volume gap, it follows that there is a maximum value of $N_{\Gamma}$, because there is a maximum number of nodes for graphs in $\mathcal{H}_{V}$, as every node carries a minimum volume $v_o$. Therefore a region of volume equal or smaller than $V$ contains at most 
\be \label{eq:volEigen}
n_V=\frac{V}{v_o}
\ee nodes. 
Equation \eqref{eq:vmin} bounds also the number of allowed area quantum numbers, because too large a $j_{max}$ would force too large a node volume. Therefore $N_{\{j_l\}}$ is also finite. Finally, since the dimension of the space of the intertwiners at each node is finite and bounded by the value of spins, it follows that also the number $N_{\{v_n\}}$ of individual volume quantum numbers is bounded.   Then \eqref{eq:boundDim} shows immediately that the dimension of $\mathcal{H}_V$ is finite. Let us bound it explicitely.
 
We start by the number of graphs.  The number of nodes must be smaller than $n_V$, given in \eqref{eq:volEigen}. The number $N_\Gamma$ of 4-valent graphs with $n$ nodes is bounded by  
\begin{equation} \label{1}
N_{\Gamma} \leq n^{4 n}  
\end{equation}
because each node can be connected to other $n-1$ four times. 

Equation \eqref{eq:vmin} bounds the spins. Since we must have $V\ge v_{gap}$, we must also have 
\be 
j \le j_{max} \leq 32\frac{V^2}{\ell_{\rm P}^6 {{\gamma^{3}}}}=\frac1{2}n_V^2
\ee 
In a graph with $n$ nodes there are at most $4n$ links (the worst case being all boundary links), and therefore there are at most $\left(2 j_{max} +1\right)^{4n}$ spin assignments, or, in the large $j$ limit, $\left(2 j_{max}\right)^{4n}$.  That is 
\be\label{2}
N_{\{j_l\}}\le \left(2 j_{max}\right)^{4n} \le n^{8n}
\ee

Finally, the dimension of the intertwiner space at each node is bounded by the areas associated to that node:
\begin{align*}
& \dim \mathcal{K}_{j_1,j_2,j_3,j_4}  = \\
 & = \dim \;  {\tt Inv}_{SU(2)}  \left( \mathcal{H}_{j_1}  \otimes  \mathcal{H}_{j_2} \otimes \mathcal{H}_{j_3} \otimes \mathcal{H}_{j_4} \right)  \\
 & = {\tt min}  \left(  j_1+j_2,j_3+j_4\right)  - {\tt max} \left((j_1-j_2),(j_3-j_4)\right) +1 \\
 & \leq 2 \; {\tt max}(j_{l \in n}) +1   \leq 4 \; {\tt max}(j_{l \in n}) 
\end{align*}
with the last step following from ${\tt max}(j_{l \in n})  \geq 1/2$. 
Thus on a graph with $n$ nodes, the maximum number of combinations of eigenvalues is limited by:
\begin{equation}\label{3}
N_{\{v_n\}} \leq \left(4j_{max}\right)^{n} = 2^n n^{2n}
\end{equation}

Combining equations \eqref{eq:boundDim}, \eqref{1}, \eqref{2} and \eqref{3}, we have an explicit bound on the dimension of the space of states with volume less than $V=n_V\, v_o$:
\be
\dim \mathcal{H}_{V} \le (c n_V)^{14 n_V}  
\ee
where $c$ is a number.  For large $n_V$ we can write 
\be
S_V\equiv \log \, \dim \mathcal{H}_{V}  \le  14\, n_V
\log n_V
\ee
which is the entropy associated to this Hilbert space. Explicitly
\be
S_V \le  14\; \frac{V}{v_{o}}\log \frac{V}{v_o} \sim  V \log V.
\ee
In the large volume limit, when the eigenvalues become increasingly dense, this corresponds to a density of states $\nu(V)\equiv d(\dim \mathcal{H}_{V})/dV$ similarly bounded
\be
\nu(V) <   14\left[\log\left(\frac{cV}{v_0}\right) + 1 \right]\left(\frac{cV}{v_0}\right)^{14 \frac{V}{v_0}} .
\ee
\medskip

\section{Von Neumann proper volume entropy} \label{sec:3}

In the previous section, we have observed that the dimension of the space of quantum states (with four-valent, finite-volume nodes) with total volume less than $V$ is finite.   This result implies that there is a finite von Neumann volume entropy associated to statistical states describing volume measurements.  

The simplest possibility is to consider the micro-canonical ensemble describing the volume measurement of a region of space.  That is, we take Volume to be a macroscopic (or  thermodynamic, ``coarse grained") variable, and we write the corresponding statistical microstate that maximizes entropy. If the measured volume is in the interval $I_V=[V-\delta V,V]$, with small $\delta V$, then the corresponding micro-canonical state is simply 
\begin{equation}
\rho =  \frac{\mathcal{P}_{V, \delta V}}{\dim\mathcal{H}_{V, \delta V}}.
\label{marios}
\end{equation}    
where $\mathcal{P}_{V, \delta V}$ is the projector on 
\begin{equation} \label{eq:HkDef}
\mathcal{H}_{V, \delta V} \equiv Span\{|\Gamma, j_l , v_n>\;\; :\; v \in I_V\}.  
\end{equation}
namely the span of the eigenspaces of eigenvalues of the volume that are in $I_V$.  Explicitly, the projector can be written in the form 
\begin{equation}
\mathcal{P}_{V, \delta V} \equiv \;\sum_{v\in I_V} \;|\Gamma, j_l , v_n> <\Gamma , j_l, v_n|  
\end{equation}
The von Neumann entropy of \eqref{marios} is 
\be
S = - Tr[\rho\log{\rho}]  =\log{\dim\mathcal{H}_{V, \delta V}} < S_V\sim V\log V.
\ee

It is  interesting to consider  also a more generic state where $\rho\sim p(V)$, for an arbitrary distribution $p(V)$ of probabilities of measuring a given volume eigenstate with volume $V$. 
For this state, the probability distribution of finding the value $V$ in a volume measurement is 
\be
P(V)=\nu(V)p(V)
\ee
and the entropy can be written as the sum of two terms 
\be
S = \int dV \ \nu(V) p(V)\log(p(V)) =S_P+S_{\rm Volume}
\ee
where the first
\be
S_P =- \int dV \ P(V)\log(P(V)) 
\ee
is just the entropy due to the spread in the outcomes of volume measurements, while the second
\be
S_{\rm Volume} \equiv S-S_P = \int dV \ P(V)\log(\nu(V)) 
\ee
can be seen as as a \emph{proper} volume entropy.  The bound found in the previous section on $\nu(V)$, which indicates that $\log (\nu(V))$ grows less that $V^2$, shows that this proper volume entropy is finite for any distribution $P(V)$ whose variance is finite. $S_{\rm Volume}$ can be viewed as the irreducible entropy associated to any volume measurement. 

\section{Lower bound} \label{sec:5}

Let us now bound the dimension of $\mathcal{H}_V$ from below.  The crucial step for this  is to notice the existence of a maximum $\delta V$ in the spacing between the eigenvalues of the operator $\hat V_n$.  For instance, if we take a node between two large spins $j$ and two $\frac12$ spins, the volume eigenvalues have decreasing spacing, with maximum spacing for the lowest eigenvalues, of the order $v_o$. Disregarding irrelevant small numerical factors, let's take $v_o$ as the maximal spacing.  

Given a volume $V$ let, as before, $n_V=V/v_0$ and consider spin networks with total volume in the interval $I_n=[(n-1)v_o,nv_o]$. Let $N_m$ be the number of spin networks with $m$ nodes that have the total volume $v$ in the interval $I_n$.  For $m=1$, there is at least one such spin network, because of the minimal spacing.   For $m=2$, the volume $v$ must be split between the two nodes: $v=v_1+v_2$. This can be done in at least $n-1$ ways, with 
$v_1\in I_p$ and $v_2\in I_{n-p}$ and $p$ running from $1$ to $n-1$.  This possibility is guaranteed again by the existence of the maximal spacing. In general, for $m$ nodes, there are  
\begin{equation}
N_{n,m} =  {{n-1}\choose{m-1}}
\end{equation} 
different ways of splitting the total volume among nodes. This is the number of \emph{compositions} of $n$ in $m$ subsets. Finally, the number $m$ of nodes can vary between $1$ and the maximum $n$, giving a total number of possible states larger than 
\begin{equation}
N_n = \sum_{m=1}^{n}N_{n,m}  = \sum_{m=1}^{n}{{n-1}\choose{m-1}} = 2^{n-1}.  
\end{equation}
From which it follows that 
\begin{equation}
\dim \mathcal{H}_V  \geq  2^{n_V-1}.  
\end{equation}

Can all these states be realised by inequivalent spin networks, with suitable choices of the graph and the spins? To show that this is the case, it is sufficient to recognise that there exists at least one (however peculiar) example of spin network for each sequence of $v_n$: given an arbitrary sequence of $v_n$, we can always construct a graph formed by a single one dimensional chain of nodes, each (except the two ends) with two links connecting to the adjacent nodes in the chain and two links in the boundary. All these spin networks exist and are non-equivalent to one another. 

To be clear, there will be of course many more graphs for any given volume configuration, but, given that we are looking for a lower bound, it is sufficient to find one example for every sequence of $v_n$. A graph composed by a one dimensional chain of nodes, each (except the two ends) with two 
links connecting to the adjacent nodes in the chain and two links in the boundary, leaves complete freedom to the assignment of volume to every node. Thus, for the purposes of extracting a lower bound we must count at least one graph for each sequence $v_n$.

Therefore we have shown that there are at least $2^{n_V-1}$ states with volume between $V - v_o$ and $V$. In the large volume limit we can write 
\begin{equation}
\dim \mathcal{H}_V \geq  2^{n_V} = 2^{\frac{V}{v_o}}.  
\end{equation}
so that the entropy satisfies 
\begin{equation}
 cV \le S \le c'V \log V. 
\end{equation}
with $c$ and $c'$ constants.

\section{Discussion} \label{sec:7}

Geometrical entropy associated to \emph{surfaces} of given Area plays a large role in the current discussions of the nature of spacetime \cite{Wheeler:1990uq,Wheeler:1991fs,JacobsonThermodynamicsSpacetimeEinstein1995,RovelliBlackHoleEntropy1996,Bousso, Bianchiarchitecturespacetimegeometry2014,Marolf}.  
Here we have shown that, under suitable conditions, it is also possible to compute the von Neumann entropy associated to measurements of the Volume of a \emph{region} of space.   We have not discussed possible physical roles played by this entropy. A number of comments are in order:
\begin{enumerate}[(i)]
\item Since in the classical low energy limit volume and area are related by  $V\sim A^{\frac32}$, the Volume entropy we have considered $S_V\sim V\log V\sim A^{\frac32} \log A$ may exceed the Bekenstein bound $S<S_A\sim A$.  Volume entropy is accessible only by being in the bulk, and not necessarily from the outside, therefore it does not violate the versions of the Bekenstein bound that only refer to external observables. 
\item The result presented above depends on the restriction of $H_{\rm\emph{diff}}$ to four-valent states.  We recall that the discussion is currently open in the literature on which of the two theories, with or without this restriction, is physically more interesting, with good arguments on both sides.  However, it might be possible to extend the results presented here to the case of higher-valent graphs. Indeed, there is some evidence that there is a volume gap in higher-valent cases too, see for instance \cite{HaggardPentahedralvolumechaos2013a}. 
The effect of zero-volume nodes on the Volume entropy will be discussed elsewhere. 

\item It has been recently pointed out that the interior of an old black hole contains surfaces with large volume \cite{ChristodoulouHowbigblack2015a,
ChristodoulouVolumeoldblack2016b} and that the large volume inside black holes can play an important role in the information paradox \cite{PerezNofirewallsquantum2015,AbhayILQGS}.   The results presented here may serve to quantify the corresponding interior entropy. 

\item A notion of entropy associated to the volume of space might perhaps provide an alternative to Penrose's Weyl curvature hypothesis \cite{PenroseBigBang}. For the second principle of thermodynamics to hold, the initial state of the universe must have had low entropy. Penrose proposes to address this low entropy by taking into consideration the entropy associated to gravitational degrees of freedom. His hypothesis is that the degrees of freedom which have been activated to bring the increase in entropy from the initial state are the ones associated to the Weyl curvature tensor, which in his hypothesis was null in the initial state of the universe. A definition of the bulk entropy of space, which, as would be expected, grows with the volume, could perhaps perform the same role as the Weyl curvature degrees of freedom do in Penrose's hypothesis: the universe had a much smaller volume close to its initial state, so the total available entropy was low - regardless of the matter entropy content - and has increased since then, just because for a space of larger volume we have a greater number of states describing its geometry. 
\item We close with a speculative remark. Does the fact that entropy is large for larger volumes imply the existence of an entropic force driving to larger volumes?  That is, could there be a statistical bias for transitions to geometries of greater volume? Generically, the growth of the phase space volume is a driving force in the evolution of a system: in a transition process, we sum over  \emph{out} states, so more available states for a given outcome imply greater probability of that outcome. A full discussion of this point requires the dynamics of the theory to be explicitly taken into account, and we postpone it for future work.
\end{enumerate}

\section*{Acknowledgments}
MC and VA thank Thibaut Josset and Ilya Vilenski for useful discussions. MC acknowledges support from the Educational Grants Scheme of the A.G.Leventis Foundation for the academic years 2013-2014, 2014-2015, 2015-2016, as well as from the Samy Maroun Center for Time, Space and the Quantum. VA acknowledges support from the Sapienza University of Rome.

\vfill

\bibliography{VolumeEntropy.bbl}

\providecommand{\href}[2]{#2}\begingroup\raggedright\begin{thebibliography}{10}

\bibitem{BekensteinBlackHolesEntropy1973a}
J.~D. Bekenstein, {\it Black {{Holes}} and {{Entropy}}},  Physical Review D
  {\bf 7} (Apr., 1973) 2333--2346.

\bibitem{Bardeenfourlawsblack1973}
J.~M. Bardeen, B.~Carter and S.~W. Hawking, {\it The four laws of black hole
  mechanics},  Communications in Mathematical Physics {\bf 31} (June, 1973)
  161--170.

\bibitem{HawkingParticlecreationblack1976}
S.~W. Hawking, {\it Particle creation by black holes},  Communications in
  Mathematical Physics {\bf 46} (June, 1976) 206--206.

\bibitem{JacobsonThermodynamicsSpacetimeEinstein1995}
T.~Jacobson, {\it Thermodynamics of {{Spacetime}}: {{The Einstein Equation}} of
  {{State}}},  Physical Review Letters {\bf 75} (Aug., 1995) 1260--1263.

\bibitem{ChircoSpacetimethermodynamicshidden2014}
G.~Chirco, H.~M. Haggard, A.~Riello and C.~Rovelli, {\it Spacetime
  thermodynamics without hidden degrees of freedom},  Physical Review D {\bf
  90} (Aug., 2014).

\bibitem{BianchiEntanglement}
E.~Bianchi, {\it {Horizon entanglement entropy and universality of the graviton coupling}},
  \href{https://arxiv.org/abs/1211.0522}{{\tt arXiv:1211.0522}} (2012). 

\bibitem{Bianchiarchitecturespacetimegeometry2014}
E.~Bianchi and R.~C. Myers, {\it On the architecture of spacetime geometry},
  Classical and Quantum Gravity {\bf 31} (Nov., 2014) 214002.

\bibitem{RovelliBlackHoleEntropy1996}
C.~Rovelli, {\it Black {{Hole Entropy}} from {{Loop Quantum Gravity}}},
  Physical Review Letters {\bf 77} (Oct., 1996) 3288--3291.

\bibitem{PerezBarbero}
F.~Barbero, A.~Perez {\em Quantum Geometry and Black Holes}.
\newblock {Loop Quantum Gravity: The First 30 Years, WSP, 241-279},
  2017.

\bibitem{AbhayILQGS}
A.~Ashtekar, ``\emph{The Issue of Information Loss: Current Status}.'' Talk
  given at the International Loop Quantum Gravity Seminar. Audio archived
  online, 2016.
  
\bibitem{Rovelli17}. C.~Rovelli, ``\emph{Black holes have more states than those giving the Bekenstein-Hawking entropy: a simple argument}.", arXiv:1710.00218.

\bibitem{ChristodoulouHowbigblack2015a}
M.~Christodoulou and C.~Rovelli, {\it How big is a black hole?},  Physical
  Review D {\bf 91} (Mar., 2015).

\bibitem{ChristodoulouVolumeoldblack2016b}
M.~Christodoulou and T.~De~Lorenzo, {\it Volume inside old black holes},
  Physical Review D {\bf 94} (Nov., 2016).

\bibitem{PerezNofirewallsquantum2015}
A.~Perez, {\it No firewalls in quantum gravity: The role of discreteness of
  quantum geometry in resolving the information loss paradox},  Classical and
  Quantum Gravity {\bf 32} (Apr., 2015) 084001.


\bibitem{MeissnerEigenvaluesvolumeoperator2006}
K.~A. Meissner, {\it Eigenvalues of the volume operator in loop quantum
  gravity},  Classical and Quantum Gravity {\bf 23} (Feb., 2006) 617--625.

\bibitem{RovelliSingleparticlequantum2010a}
C. ~Rovelli,F.~ Vidotto {\it Single particle in quantum gravity and Braunstein-Ghosh-Severini entropy of a spin network},  Physical Review D {\bf 81} (2010). 


\bibitem{LivineBulkEntropyLoop2008}
E.~R. Livine and D.~R. Terno, {\it Bulk Entropy in Loop Quantum Gravity},  Nuclear Physics B {\bf 794} (2008) 138--153.

\bibitem{RovelliCovariantloopquantum2015}
C.~Rovelli and F.~Vidotto, {\em Covariant Loop Quantum Gravity: An Elementary
  Introduction to Quantum Gravity and Spinfoam Theory}.
\newblock {Cambridge University Press}, Cambridge, United Kingdom ; New York,
  2015.

\bibitem{QuantumGravityRovelli}
C.~Rovelli, {\em Quantum Gravity}.
\newblock {Cambridge Monographs on Mathematical Physics},
  2014.

\bibitem{ThiemannLecturesLoopQuantum2003}
T.~Thiemann, {\em Lectures on Loop Quantum Gravity}.
\newblock {arXiv:gr-qc/0210094},
  2003.

\bibitem{ThiemannModernCanonicalQuantum2007}
T.~Thiemann, {\em Modern Canonical Quantum General Relativity}.
\newblock {Cambridge University Press},
  2007.


\bibitem{BrunnemannPropertiesvolumeoperator2008a}
J.~Brunnemann and D.~Rideout, {\it Properties of the volume operator in loop
  quantum gravity: {{I}}. {{Results}}},  Classical and Quantum Gravity {\bf 25}
  (Mar., 2008) 065001.

\bibitem{BrunnemannPropertiesvolumeoperator2008b}
J.~Brunnemann and D.~Rideout, {\it Properties of the volume operator in loop
  quantum gravity: {{II}}. {{Detailed}} presentation},  Classical and Quantum
  Gravity {\bf 25} (Mar., 2008) 065002.

\bibitem{BrunnemannSimplificationspectralanalysis2006}
J.~Brunnemann and T.~Thiemann, {\it Simplification of the spectral analysis of
  the volume operator in loop quantum gravity},  Classical and Quantum Gravity
  {\bf 23} (Feb., 2006) 1289--1346.

\bibitem{BianchiDiscretenessVolumeSpace2011a}
E.~Bianchi and H.~M. Haggard, {\it Discreteness of the {{Volume}} of {{Space}}
  from {{Bohr}}-{{Sommerfeld Quantization}}},  Physical Review Letters {\bf
  107} (July, 2011).

\bibitem{HaggardPentahedralvolumechaos2013a}
H.~M. Haggard, {\it Pentahedral volume, chaos, and quantum gravity},  Physical
  Review D {\bf 87} (Feb., 2013).


\bibitem{Wheeler:1990uq}  J.A.~Wheeler, in ``Complexity, Entropy, and the Physics of Information" (ed. Zurek, W. H.) 3Ð28 (Addison-Wesley, 1990).

\bibitem{Wheeler:1991fs} J.A.~Wheeler, \emph{It from bit}. in Moscow 1991, Proceedings, Sakharov memorial lectures in physics 2, 751 (1991).

\bibitem{Bousso} R.~Bousso,  \emph{The holographic principle}. Reviews of Modern Physics. 74  825Ð874. arXiv:hep-th/0203101 (2002) 

\bibitem{Marolf} D.~ Marolf, \emph{The Black Hole information problem: past, present, and future}. Reports Prog. Phys. 80, 092001 (2017).

\bibitem{PenroseBigBang}
R.~Penrose, {\it {Before the big bang: An outrageous new perspective and its implications for particle physics
}}, Conf.Proc. C060626 (2006) 2759-2767



\end{thebibliography}\endgroup

\end{document}